# A microstructural analysis of 2D halide perovskites: Stability and functionality

**Susmita Bhattacharya\*, \$,1, Goutam Kumar Chandra\*2, P. Predeep2**

[1] Department of Physics, RGUKT Nuzvid, Nuzvid, Andhra Pradesh, India

[2] Department of Physics, NIT Calicut, Kozhikode, Kerala, India

[$] Present address: Radiation Laboratory, University of Notre Dame, Notre Dame, Indiana 46556, USA

**\* Correspondence:**
Dr. Susmita Bhattacharya
bhattacharyasusmita7@gmail.com
Dr. Goutam Kumar Chandra
goutam@nitc.ac.in

**Keywords: halide perovskite, 2D material, layered perovskite, device, solar cell**

## ABSTRACT

Recent observations indicated that the photoelectric conversion properties of perovskite materials are intimately related to the presence of superlattice structures and other unusual nanoscale features in them. The low dimensional or mixed dimensional halide perovskite family are found to be more efficient materials for device application compared to 3-dimensional halide perovskites. The emergence of perovskite solar cell has revolutionized the solar cell industry because of their flexible architecture and rapidly increased efficiency. Tuning the dielectric constant, charge separation are the main objective in designing a photovoltaic device that can be explored using 2-dimensional perovskite family. Thus, revisiting the fundamental properties of perovskite crystals could reveal further possibilities for recognizing these improvements towards device functionality. In this context, this review discusses the material properties of 2-dimensional halide perovskite and related optoelectronic devices aiming particularly for solar cell application.

## 1    INTRODUCTION

As witnessed in the past decade, the halide perovskite family has emerged as a high-performance photovoltaic material. This reappearance has involved extensive development of the three-dimensional (3D) perovskites, specifically $CH_3NH_3PbI_3$.The crystallographic formula $ABX_3$ of halide perovskite (HP) materials comprised of A, organic cation such as methylammonium (MA): $CH_3NH_2^+$) or alkali cation or mixing of both, B the metal cation ($Pb^{+2}$ or $Sn^{+2}$), and X the halide anions ($X = I^-$, $Br^-$, or $Cl^-$). HPs are mainly classified into two categories depending on the occupancy of A-site whether an organic-molecular cation or an elemental cation. The catergories are organic-inorganic hybrid HPs and all-inorganic HPs, respectively. In hybrid HPs, the organic cations occupying A-site of the perovskite structure are usually non-symmetric and inclined to rotation at room temperature and above (Chen et al., 2015). This extraordinary phenomena has been hardly observed in other materials. As a result, hybrid HPs have the interconnected $[BX_6]^4$ octahedra continued in an ordered manner that form the crystal structure with local disorder located within the $[BX_6]^4$ framework. At room temperature, the most-reviewed hybrid HP, $CH_3NH_3PbI_3$ or ($MAPbI_3$), the reorientation of polar $MA^+$ ion are found in $[PbI_6]^4$ octahedral (between the faces, corners, or edges) having a residence time of ~14 ps (Leguy et al., 2015). At lower temperatures, the rotational dynamics in HPs are slowed down (Wu et al., 2017) following their dielectric response. HPs can have different tunable chemical compositions, such as $A_2BX_4$, $ABX_4$, $A_3BX_9$, well known as Ruddlesden-Popper (RP) phases (Yu et al., 2017). The defects in the crystal, crystallographic alignments, surfaces, grain boundaries, and interfaces in HPs can be customized preferentially. These preferential nano- and micro-structures can be addressed using in-depth conventional microscopic characterization methods such as atomic force microscopy (AFM), scanning electron microscopy (SEM) and transmission electron microscopy (TEM). As an example, consider the case of a perovskite solar cell (PSC). A PSC includes a hybrid organic-inorganic or all-inorganic perovskite structured compound as the light-harvesting active layer. The compositional variation of the material and its dependency on phase transition temperature are related to electrical and ionic properties of the active layer (Correa-Baena et al., 2017, Zhou et al., 2019, Lee et al., 2012). Thus we can infer that measuring a minute variation of structural property and optimizing the same will be a quality control measure for optoelectronic devices.

The large exciton binding energy due to the reduced dielectric screening and quantum confinement are the most fascinating optoelectronic properties of two-dimensional (2D) materials. 2D HPs with its tuneable photophysical properties and superior chemical stability compared to their 3D counterpart (Ding et al 2018, Zheng et al 2018, Dong et al 2019) have created a great influence on the semiconductor research and technology fields. 2D HPs obey stereochemical rules that determine the growth of the perovskite sheets, (i)corner-sharing, (ii)edge-sharing, and (iii)face-sharing- octahedral connections has been experimentally observed. Here the characteristics for size restriction for A′ site is determined by the suitability of a cation spacer whether a) its net +ve charge at the perovskite attaching point and substitution amount (b) capacity of hydrogen bonding (c) flexible stereochemical arrangement, (d) ability of space filling (Mao et al 2019). Most cations are found to have corner sharing 2D structure but sometime the cation-imposed configuration strain in 2D network that stabilizes edge-sharing or face sharing networks (Kamminga et al 2016). In 2D perovskites, extra larger organic cations (L) are inserted as spacers, separating the inorganic metal halide octahedra layers to form quantum well superlattices that is different from conventional 3D HP. Here introducing hydrophobic spacer cations, the ionic lattice of inorganic octahedrons are effectively isolated from the ambient water molecules. The newly introduced additional spacing cations and asymmetric lattice structures provide tunable intrinsic physical features, such as dielectric constant, optical band gap, and exciton binding energy, (Blancon et al., 2017). As a result, the photophysical behaviors like exciton dynamics (Proppe et al., 2019), charge carrier transport (Tsai et al., 2018), and electron−phonon coupling (Straus et al., 2018) that strongly influence the performance of optoelectronic devices (light emitting diode (LED), solar cell(SC)) can also be modulated accordingly (Zheng et al., 2018). In a recent report, the mechanical properties of $(C_6H_5CH_2NH_3)_2PbCl_4$ are found to be anisotropic, and the organic components and van der Waals interactions between layers play a significant role in the structural stability of the 2D structure (Gao et al. 2020). They found that the substituting the organic parts with rigid and multifunctional organic components with Pb will improved stability and carrier mobility of the PSC absorber layer.

The interaction of electromagnetic fields with photo- and electro-active materials is the heart of energy conversion research. Graphene, hexagonal boron nitride, transition metal dichalcogenides, MXenes *etc.* other classical 2D materials are being widely studied due to their fascinating properties such as high electrical conductivity, low density, large surface area, tunable electric and magnetic properties (Novoselov et al 2004, Mas-Ballesté et al 2011, Jiang et al 2020). Pristine graphene is a zero-bandgap semimetal. while RGO and GO are a semiconductor and an insulator, respectively. 2D TMDs display semiconducting nature(Rao et al., 2009, Ferrari et al., 2015). MXenes sometimes behaves as metals, semiconductors, superconductors, topological insulators, and, most importantly, half-metallic characteristics (Naguib et al., 2011, Naguib et al., 2012, Anasori et al., 2015). graphene and related two-dimensional (2D) materials propose extraordinary advancement towards in device performance even at the atomic scale. A compromised combination of these 2D materials with silicon chips promotes a massively enhanced potential compatible for on silicon technology. Low cost, simple and inexpensive layered two-dimensional (2D) halide perovskites are a novel class of materials with outstanding layer-dependent tunable optical properties, generation of room temperature stable excitons, and modified exciton binding energy originating from the quantum and dielectric confinement effects. They also offer the improved environmental stability and photostability needed for the light absorbing layer photovoltaic devices. The enormous potential of this thin film SC technology is expected to become a low-cost commercial alternative in the near future to the presently available silicon photovoltaic. This manuscript evolves as follows: introduction to material properties, highlighting 2D perovskite structural aspects. Next, the characterization that identifies exotic material property of 2D HP has been discussed. Finally, the microstructural characteristics of 2D HPs that leads to promising potential applications as high-performance optoelectronic devices especially SC has been pointed out.

## 2 STRUCTURAL ASPECTS OF 2D HP: ENGINEERED ENERGY LANDSCAPE

Microscopically, layered HPs contain of mono or few thin atomic layers of HP (Gao et al., 2018). Atomically (morphologically) thin HPs or crystallographically layered HPs are considered as 2D/quasi-2D (q-2D) HPs. Dimensional confinement upon crystallization of HPs using the long-chain organic ligands-templated growth, (Tyagi et al., 2015, Liu et al., 2017, Sheng et al., 2018) or self-templated growth, (Ha et al 2014, Liu et al 2016, Niu et al. 2016) preserving $ABX_3$ formula results morphological 2D/q-2D HPs. Using bulky organic separators and by slicing 3D HPs along $\langle 100 \rangle$, $\langle 110 \rangle$, and $\langle 111 \rangle$ crystal plane three different layered crystallographic 2D/q-2D HPs are obtained. As a consequence, the chemical formulae of 2D HPs modified methodically according to the number of layers and crystal plane orientation. The chemical formula of $A'_2A_{n-1}B_nX_{3n+1}$ specifically belongs to $\langle 100 \rangle$, $A'_2A_mB_mX_{3m+2}$ $\langle 110 \rangle$, and $A'_2A_{q-1} B_qX_{3q+3}$ $\langle 111 \rangle$ oriented HP family respectively. The $\langle 100 \rangle$ oriented layered HP family, owing to their high acceptance to many distinct organic and inorganic components, is the most extensively investigated among them. The HP family oriented along $\langle 100 \rangle$ direction are allocated further into two sub-categories, RP and Dion–Jacobson (DJ) (Ortiz-Cervantes et al. 2019) type. However, the DJ HPs are not often studied and applied in practical devices, thus less popular in device field (Mao et al. 2018). The name of RP HPs derived from their resemblance in crystal with inorganic $Sr_3TiO_7$ RP perovskites (Ruddlesden et al. 1958). To emphasis on the comparison in crystal structure of HPs we will focus our discussion to a schematic. Figure 1 (a) describes schematic of the 3D structure and the [001] projections of conventional inorganic $CsPbBr_3$ cubic and orthorhombic perovskite phases and RP phases as labelled in ref (Yu et al., 2017). Here, the material $Cs_{n+1}Pb_nBr_{3n+1}$ or $CsBr(CsPbBr_3)_n$, is found to be composed of "n" layers of $CsPbBr_3$ unit cells separated by an extra CsBr layer. The RP phases are considered as a repeatedly integral extrinsic stacking error, designed by shifting two neighbouring $CsPbBr_3$ units by an in-plane lattice vector (1/2 1/2) compared to each other. Figure 1(b) compares between experimental and simulated scanning TEM (STEM) images of typical domains in 2D nanosheets of $CsPbBr_3$. The three panels compares perovskite I: 6-layer perovskite phase (upper panel), 1-layer n = 2 RP phase (middle panel), and perovskite II: 6-layer perovskite phase with $-PbBr_3$ termination (bottom layer). From the figure it is clearly observed that RP domains and perovskite domains are well-connected in 2D nanosheets owing to their crystal structures similarity, and all of these domain boundaries aligned in the ⟨100⟩ directions at atomic-scale. The details of the STEM mode of operation/characterisation will be discussed in the next section.

For clarity, the general formula for 2D hybrid RP perovskite can be rewritten as $(RNH_3)_2MX_4$ or $(NH_3RNH_3)MX_4$, where M is metal cation including $Pb^{+2}$, $Cu^{+2}$, and $Sn^{+2}$, R: prototype organic molecule e.g. phenethylammonium (PEA) or butylammonium (BA) (Mitzi et al 2001). In hybrid RP perovskites, a extraordinary ability to regulate the thickness of the inorganic layers are found, which is a highly needed characteristics for photovoltaic (PV) applications, specifically SCs. As a consequence, RP perovskites are found to be the most used and studied. For the construction of interconnected few layers of corner-sharing $M-X_6$ octahedra in q-2D HPs, the short-chain inorganic or organic cations are required. The value of "n" in the general formulae considered here is the number of layers in 2D and q-2D HPs. For n = 1, we get the 2D HPs. q-2D HPs are obtained for n ≥ 2. The intrinsically insulating large organic layers induce multiple quantum well structures (Chen et al 2019) in between the assemblies of these organic layers with $M-X_6$ octahedron layers. As a result, the 2D and q-2D RP HPs are found to exhibit electronic and quantum confinement even in thin films (Ding et al., 2019, Spanopoulos et al., 2019). A comparison between 2D and 3D network of perovskite and the modified band structure has been described in ref (Straus et al 2018). Figure 2 (A) describes projection of the 3D hybrid perovskite structure. It highlights an inorganic linkage of corner sharing metal halide octahedral described by red colour, while the interstitial organic cations are marked with blue, black colour. Green colour region emphasise the limit on cation size. A projection down the *c*-axis of the 2D hybrid perovskite is schematically depicted in Figure 2(B). It shows the successive arrangement of organic layer as shown in blue and black colour and red inorganic layers that shares the corners. The thickness of inorganic and organic layers are marked as *d* and *L* respectively, for n=1. Here the green region emphasis the restriction on the area of cross section but not on the of organic cation length. The energy band diagram of 2D structures, valence band (VB), conduction band (CB), electronic bandgap ($E$g), and optical or excitonic bandgap $E_x$ (blue colour) are labelled in the Figure 2(C). The organic framework (with lager HOMO-LUMO gap) has a dielectric constant $\epsilon_2$ while the inorganic slabs have a $\epsilon_1$ ($\epsilon_1 > \epsilon_2$). A dielectric confinement effect is originated by the alternation of low dielectric constant of the organic moiety and a high dielectric constant of the $Pb^{+2}$ halide layers (Ishihara et al., 1989, Koh et al., 2017). In 2D perovskites due to a dielectric confinement effect in the layers, the electrons are strongly attracted to the holes. They attribute higher exciton binding energies for 2D compared to their 3D counterpart.

For SCs application, the exciton binding energy should be as small as possible. The lead HPs with n ≥ 3, have bandgap, $E_b$ comparable to 3D perovskites the dielectric constant increases largely with *n*, are able to perform decently as SC absorbers (Pazoki et al., 2018). The bandgap is one of the most crucial parameters in SC applications. It determines the maximum theoretical efficiency for the corresponding single p-n junction device according to the Shockley-Queisser limit (Rühle et al., 2016). A bandgap of 1.34 eV results in a power conversion efficiency (PCE) of about 33.5%, (Rühle et al., 2016, Slavney et al., 2017) while materials with bandgaps >1.9 eV are found to be the smart choice for tandem SCs. The bandgap for 2D HP is typically determined by the composition and thickness of the inorganic layers. The bulkier halides lessen the bandgap. For n = 1, 2D perovskites with formula $(PEA)_2PbX_4$, have bandgaps of about 3.8, 3.0 and 2.5eV for Cl⁻, Br⁻ and I⁻respectively (Kitazawa et al., 1997). The heavier, less electronegative halides have more delocalized electronic density and it better overlap with that of the metal atoms (Slavney et al., 2017). Therefore, halide substitutions can be a method for systematically engineered the bandgap of 2D hybrid HP by adjusting their composition of the chemical constituents. The procedure allows broad absorption across visible and near-infrared region of wavelengths. The other way of the bandgap and photoluminescence (PL) wavelength modification is by changing the *n* value. With increase in *n* , the bandgap becomes smaller, and tends towards the bandgap value of 3D perovskite as *n* approaches to ∝, and the PL peak shifts asymptotically towards lower energies with increase in *n* (Stoumpos et al., 2017). Again, the coefficient of absorption (COA) is directly proportional to the strength of electronic dipole transitions. In 2D-perovskites, light absorption is lesser due to a low density of metal/halogen atoms causing a lower absorption cross-section. COA increases rapidly as the *n* value increases. 2D perovskites with n ≥ 3, are comparable to 3D perovskites, have high enough COA can harvest most of the incoming radiation. Following the trend, a general scaling law was proposed to determine the binding energy of Wannier-Mott excitons in perovskite quantum wells of arbitrary layer thickness (Blancon et al., 2018). In another approach (Mitzi et al., 1995), modified orientation of the conducting perovskite layers through choice of organic cation has been achieved. Interestingly, they found conducting layered organic-inorganic halide containing <110> oriented perovskite sheet. Here it must be mentioned that surface modification also plays a critical role in device engineering. It was also observed that the layer edge electrons which are not related to the surface charging effect; but associated with the local energy states corresponding to of the edge electronic structure (Wang  et al., 2019). Therefore, in the next section the focus of discussion is on the characterisation of engineered 2D HP energy landscape and extrapolation of these specific features towards device application.

# 3 MICROSTRUCTURAL CHARACTERISATION 2D HP: IMPROVEMENT OF DEVICE FUNCTIONALITY

## 3.1 Crystallographic Characterisation

Conductivities and carrier mobilities of layered 2D perovskite depends on the crystallographic direction. Generally the conductivity and mobility are found to be significantly improved when measured along the planes of the inorganic layers. TEM and x-ray crystallographic technique helps to understand the device crystalline feature needed for improved device performance. A comparative structural study of 3D and 2D HPs has been followed in this section.

The specific interactions between an electron beam with high energy and a thin electron-transparent sample are captured as the contrast of TEM images [already discussed in Figure 1 of earlier section]. High resolution (resolution ~ atomic scale) imaging, can be achieved in STEM mode, with few detectors to generate the images (example BF: bright field, ADF: annular dark field, and HAADF: high-angle annular dark field). The crystallographic information is extracted from electron diffraction patterns (EDPs) of the area of interest in the specimen understudy. Due to the modification of Bragg conditions in thin films, EDPs suffer from 2D imprecise crystallographic data. Energy dispersive spectroscopy (EDS) and electron energy loss spectroscopy (EELS) gives the site-specific chemical composition information. EELS allows to gather evidences about a range of sample- related aspects, such as thickness, chemical surrounding, bonding, bandgap etc. (Yu et al., 2017, Virdi et al., 2016). On the other hand, novel EDS detectors allows efficient collection of specimen compositional data even at the atomistic scale. In a recent review the beam sensitivity of HP that guide the TEM characterization has been thoroughly discussed (Chen et al., 2020). For beam-sensitive HPs materials, leading beam-damage mechanisms involve phonon mediated heating, displacement of atoms and defect formation i.e. knock-on damage, and damage by inelastic scattering and rupture of chemical bonds i.e. radiolysis. (Egerton et al., 2004) Several approaches such as, optimizing the accelerating voltage, low-dose imaging with improved cameras can be employed to overcome degradation in beam-sensitive materials. The photon induced dynamics and polarization of $MA^+$ ions can affect the physical properties and device performance of $MAPbI_3$ HPs. Thus photon induced dynamics of $MA^+$ is a major challenge for direct imaging of organic cations in TEM. Using iterative cross-correlations-built algorithm to the filtered images, Zhang et al (Zhang et al., 2018) successfully observed the ordered nanoscale domains with off-centered $MA^+$ cations with differing orientations in $MAPbBr_3$ crystals. The $MA^+$ cations are found to exhibit configurations along the normal and parallel relative to the projection direction, forming in-plane and out-of-plane electric dipoles, respectively. Accordingly, the structural models and the corresponding simulated projected potential describes that the local deviations in the $MA^+$ orientation is responsible for ferroelectricity and/or polarization in hybrid HPs. The crystal network of hybrid HPs is much ''softer'' compared with conventional oxide perovskites. As a result, there is a high probability for local vibrations in structures, modification in structures and symmetry of the crystals. Thus to visualize the pristine structure of $CsPbBr_3$ perovskite prepared via a catalyst-free solution-phase method (Yu et al., 2016),Yu et al. employed HRTEM in addition to small-dose in-line holography. The specimen under observation exhibit an atomically layered 2D nanosheet morphology. Employing focal series, they showed the existence of dual phases of $CsPbBr_3$ cubic and orthorhombic HP within the nanosheets. TEM can also identify the misalignment between the successive grains and even describe the boundary plane. Twin boundaries in $MAPbI_3$ HP was identified by Rothmann et al. (Rothmann et al., 2017) using TEM. These twin boundaries in specimens can originate during growth, phase transition, or even under mechanical stress. As HPs, exhibit extreme chemical sensitivity, etching using a chemical solution is not a viable method for observing twin boundaries. Thus, electron back-scattered diffraction (EBSD) and TEM are essential tools for revealing twinning in HPs. Based on low-dose TEM imaging and selected area EDPs (SAEDPs), Rothmann et al. showed the presence of twinning in tetragonal $MAPbI_3$, where these boundaries appeared in the form of parallel striped bands of few hundred nm width within $MAPbI_3$ grains. Although the device performance upon incorporation of the material was not clearly described by them. However, another recent study has pointed that twin boundaries can be very effective in restricting photocarriers (Li et al., 2018). The other report (Bertolotti et al., 2017), prompts flexibility of HPs nanocrystals (NC) upon halide changes and temperature variation which has been investigated following high-resolution synchrotron X-ray total scattering data. Variations in local structure of these NCs was found to exhibits orthorhombic tilting of $PbX_6$ octahedra within locally ordered subdomains. These subdomains are linked by a 2D/3D network of twin boundaries through which the coherent arrangement of the $Pb^+$ over the whole NC is preserved. Thus, we believe that controlling the properties of twining (size, orientations, etc.) could be a viable approach for the achieving and modifying the characteristics newly designed HPs.

The grain-boundary chemistry in PSCs, intensely influence carrier recombination and potential transport of ionic and molecular species, thus affecting the optoelectronic properties and chemical stability of HPs. The grain-boundary chemistry of HPs dependents on the HP bulk elemental composition, growth conditions, the structure and the kinetics of the boundary, and careful chemical modifications. In a contemporary literature (Zong et al., 2018), triblock copolymer (Pluronic P123) functionalized MAPbI₃ HP are found to result in water-resistant HPs. The copolymer establish the interaction with HP grains via its hydrophilic tails, while its hydrophobic core is placed towards the center of the copolymer wetting film. As a consequence, a water resistant HP with higher stability was obtained.

**3.2 Morphological and conductive characterization: Scanning Probe Microscopy**

Scanning probe microscopy, including AFM, Kelvin probe force microscopy (KPFM), conductive AFM (C-AFM), has been broadly applied to explore the local physical, electromagnetic or molecular topographies at microscopic scale. AFM and KPFM are two different modes of microscopic technique that operate at constant or variable bias condition. Specifically, KPFM captures the contact potential difference (CPD) between conductive tip and the sample (Chen et al., 2019). Under an alternative voltage, the CPD is measured by using a vibrating capacitor with Kelvin method (Benstetter, Liu, Frammelsberger, and Lanza, 2017). Instead of current, the sensitivity of KPFM is improved, by recording the electrostatic force interaction between the tip and the sample. By accurate adjustment of the tip work function on the reference sample, the other materials surface work function can be determined (Shao et al., 2016). In high resolution C-AFM (Yang et al., 2009, Shin et al., 2013) the current–voltage characteristics of the sample surface could be easily captured by the mapping of the topography. A large current flow points high conductivity of sample, a strategic character of charge transport (Wang et al., 2016, Lee et al., 2018, Xu et al., 2018, Eichhorn et al 2018, Bristowe et al., 2015, Si et al., 2017). Thus, by C-AFM, the conductivity difference among dissimilar perovskite films can be captured and their transport characteristics (at grain boundaries, in grain interiors) can be distinguished with the nanoscale resolution. Even for phase pure perovskites where the compositional engineering are used to overcome difficulties in obtaining high-quality sample, ambient instability degrades the device perfomance. Lee et al, reported the fabrication of phase-pure formamidinium-lead tri-iodide perovskite films with excellent optoelectronic quality and stability. They showed that incorporation of 1.67 mol% of 2D PEA lead iodide into the precursor solution allows the formation of phase-pure formamidinium perovskite with an enhanced photoluminescence lifetime (Lee et al, 2018). Figure 3 describes the band alignment and local conductivity with 2D perovskite as described by (Lee *et al.* 2018). Figure 3(A) describes that the device schematic comprising polycrystalline film of 3D perovskite with grain boundaries containing 2D perovskite. while (B) describes the layer by layer Tauc plots and UV photoelectron spectroscopic analysis of the device band structure. The bare C-AFM images of $FAPbI_3$ and with films of 2D perovskite on ITO glass coated with $SnO_2$ are shown in Figure 3 (C, E) and (D, F) respectively. A 100-mV bias voltage was applied to get the images under the room light condition for image (C, D) and low light intensity condition for image (E, F) as provided by the setup. Topology of each film is depicted in the inset of the corresponding images. For (C, D) and (E, F), the corresponding scale bars are shown at the left and right side, respectively. It describes that under ambient light conditions, flow of current in the perovskite film with 2D perovskite was higher at/near the grain boundaries while fairly uniform flow of current was noted in the bare $FAPbI_3$ film. With light illumination, the flow of current was further enhanced at/near the grain boundaries with 2D perovskite whereas current flow in bare $FAPbI_3$ film was uniformly increased, which indicates separation of charge and collection of photo-generated electrons is more favourable at grain boundaries with 2D perovskite. On the other hand, the 2D $PEA_2PbI_4$ perovskite with aromatic rings and longer alkyl chains is projected to be more resilient to moisture. It guards the defective grain boundaries of 3D perovskite, resulting in significantly improved moisture stability of the film. Regardless of the improved stability, degraded electronic properties of the film is expected due to the poor charge carrier mobility of the 2D perovskite. Interestingly, a report by Wang et al., highlights a prominent conducting feature, observed at the layer edges between the insulating bulk regions of $(C_4H_9NH_3)_2PbI_4$ 2D perovskite single crystal. In the bulk region, the electron-hole pairs are tightly bonded. At the layer edges, additional edge state (ES) electrons are located, which control the current detected by the C-AFM measurement (Wang et al., 2019). These observations of the metal-like conducting feature at the layer edge of 2D perovskite offers a different dimension for improving the performance of the next-generation optoelectronics and innovative nanoelectronics. Piezo-response force microscopy (PFM) is an AFM-based method depending on the inverse piezoelectric effect analyzing the local electromechanical properties of piezoelectric samples. Remarkably, PFM has been widely employed to study the possible ferroelectricity in $MAPbI_3$ perovskite (Fan et al., 2015). The theoretical calculations suggested presence of ferroelectric domain in the perovskite samples that would result separation of photoexcited electron-hole pair making an internal junction. It allows reduction of recombination through segregation of charge carriers. PFM also ascertains the ferroelectric features in other 2D layered perovskite materials (Sha et al., 2019, Chen et al., 2020) useful as sensors in flexible devices, soft robotics and biomedical devices, a great influence towards next generation devices. However, ferroelectricity of $MAPbI_3$ is still under debate (Wang et al., 2017).

Here would like to mention that, other than conventional microscopic techniques microwave photoconductivity imaging (MIM) on MAPbI$_3$ has been proved to be an efficient technique for quantitative estimation of photoconductivity. A recent report highlights the nanoscale photoconductivity imaging of two MAPbI$_3$ thin films with different efficiencies by light-stimulated microwave impedance microscopy (Chu et al., 2017). Here, unlike the C-AFM technique, a direct contact between the MIM tip and the perovskite thin films is not required. The effective capacitive coupling between the MIM tip and the perovskite thin films at GHz frequencies, overcomes the difficulties of nanoscale electrical properties on samples with a capping layer which is insulating in nature.

### 3.3 Distinctive Spectroscopic Characteristics: An en route to device functionality

The two most general schemes practised to enhance the PCE of 2D and q-2D HPs-based SCs are: i) optimising the layer numbers (n) to get adequate absorption of light and charge generation, ii) refining crystallographic alignment for better charge transport to the electrode. Although the layer number can be easily modified by tuning the composition of 2D and q-2D HPs in precursors, while the controlled orientational growth remains a big challenge. Non-invasive spectroscopic approaches are the easiest to probe these distinctive characteristics of the 2D perovskite specimens compared to conventional microscopic technique as here we don't need any sample preparation. Here we will discuss, a general scaling law was proposed by Blancon et al., 2018 to determine the binding energy of Wannier-Mott excitons in perovskite quantum wells of arbitrary layer thickness. Figure 4 describes the absorption and PL i.e., optical spectroscopic feature of the 2D and q-2D HP crystals having n =1–5 for inorganic RP perovskite quantum wells following the reference (Blancon et al., 2018). It specifically highlights (A) scaling of experimentally obtained optical bandgap and (B) photoluminescence spectra for the material. This understanding provides direct insight about the photophysics of RP perovskite materials utilised in practical devices.

Generally, in 3D HPs, excited charges diffuse quicker than the radiative recombination rate. Thus, most of them are confined at the grain boundary defects. However, in 2D HPs, free charges and excitons co-exists in a small regime, enhancing the radiative combination relative to bulk specimen. Due to the surface and interface defects present in the sample, the non-radiative recombination still controls the energy transfer phenomena and the PL in 2D HPs remains comparatively low. It is evident from these results that layered HPs might not be the good choice for LEDs when one considers high brightness and effective quantum efficiency (EQE). However, owing to their excellent stability and process ability, they are considered to be finest candidates for practical LEDs (Yang et al., 2018; Chiba et al., 2018; Xu et al., 2019).

Metal HPs also show necessary optical–electrical characteristics for PV devices, such as long carrier diffusion lengths, high carrier mobility, strong and broad optical absorption, all of which support the notable PCE of the device. In general, 2D flakes of $MAPbI_3$ is a widely accepted active material for photodetectors for its large COA, high mobility of the carrier and smaller exciton binding energy (<26 meV, room temperature). 2D RP perovskites are also known to exhibit the excellent photodetection performance. In a recent report by Zhou et al. (Zhou et al., 2016) explained that the RP perovskite film displayed the layer number dependent response spectrum, which is compatible with their absorption spectrum. As the perovskite films are configured into photodetectors, better photodetection property was detected in the film with the higher n value. Here the polycrystalline nature of 2D RP perovskites might be the cause behind the observed lower photoresponse. In recent times, photodetection ability of 2D RP perovskites was observed to be considerably upgraded by the replacement of the long direct chain n-BA$^+$ with branched i-BA$^+$ using a method of hot-casting processing (Dong et al., 2018).

The functionality in the devices is mostly dependant on the grain structure of polycrystalline 2D perovskites, but in-situ, chemically specific characterization of 2D perovskite grains are currently limited. Here the polarized ultra-low-frequency Raman microspectroscopy is found to be a simplistic yet influential tool for identifying relative grain orientations in 2D perovskite thin films (Toda et al., 2020). Even from temperature-dependent PL spectra of 2D perovskites and following generalized Elliot formula, the evidence of phonon-assisted sidebands are explained by Feldstein et al. They determined the spectral linewidth of the energetically lowest excitonic transitions in view of scattering channels focussing on the emission and absorption of phonons (low-energy acoustic and higher energy optical) (Feldstein et al., 2020). The low-energy, heat-carrying acoustic phonons plays the central role for management of heat in 2D RP-supported devices, remained unexplored for a long time. A recent literature by Guo et al. discusses the generation and transmission of coherent longitudinal acoustic phonons along the 2D RPs cross-plane direction, resulting distinct characterizations of lower-bandgap refractive indices. They have demonstrated a significant drop in group velocity and propagation length of acoustic phonons in 2D RPs compared to the 3D MA lead iodide counterpart (Guo et al., 2018). The observed vibrational feature is a consequence of large acoustic impedance disparity between the successive layers of perovskite and large organic cations.

Pump-probe spectroscopy offers a feasible way to identify the carrier dynamics in the 2D perovskite materials (Giovanni et al., 2018). Transient absorption microscopy (TAM) (Williams et al., 2019) is one of the technique employed to probe diffusion of carrier and two-body recombination processes. Figure 5, as reproduced following Williams *et al.* 2019, describes the schematic of concentration distribution of quantum well that yield a gradient in excitation frequencies and enables a unidirectional flow of energy as a cascade, obtained by TAM. Figure 5 (A) describes the concentration gradient of quantum wells (QWs) throughout a film of layered perovskite. According to the idea, the glass edge of the perovskite film is highly populated with the smallest QWs and the air edge is mostly dense with thickest QWs. Figure 5 (B) Schematically highlights the diagram describing down-hill energy transfer cascade. It is observed that excitonic resonances of the quantum wells decrease with increase in thickness and the direction of electron transfer is same as that of energy transfer as compared to the holes which are moving towards the states of higher energy. For applications such as microcavity lasers, boosting of the two-body recombination processes is desirable. Here, a methodical comparison of the layered films of $(PEA)_2(MA)_{n-1}[Pb_nI_{3n+1}]$ with phase-pure single crystals disclose that diffusion is blocked by grain boundaries in the films, stimulating two-body recombination. The energy transfer rules the sub-200 ps time scale and the energy levels of the quantum wells are organised such that hole transfer may take place from the air to glass sides of the film at future times. The high-performance nanolasers made up of perovskites have attained low threshold, high quality factor and tunable wavelengths under optical excitation (Guo et al., 2018, Zhang et al., 2018, Liu et al., 2020). Their compact volume could entrap the light field in a tiny region and improve the light–matter interactions. Another interesting report establish that quantum wells of colloidal lead halide perovskites can produce fully decoupled multi quantum well (MQW) superlattices (having intralayer local exciton) with ultrathin organic quantum barriers (Jagielski et al., 2020), same as the addition of monolayer hybrid boron nitride (h-BN). The obtained result demonstrates that photonic source have emission of narrowband, high quantum yield, improved light outcoupling, and wavelength tenability useful for for nano-antenna (near field) and LED (far field). Even theoretically it was proposed (Li et al., 2020) that the use of superlattice structures is an smart approach (such as mixing of cations and partial substitution of halogens by superhalogens) for expanding the family of perovskites and obtaining excellent optoelectronic materials with improve stability.

VB of 2D perovskites is predominantly consist of halide p-orbitals hybridized with metal *s*-orbitals and a CB is of metal *p*-orbitals dominated. In lead HPs, the corresponding orbitals are $5p$ of I and $6s$ and $6p$ of Pb (Gebhardt et al., 2017, Even et al., 2013, Even et al., 2012). In 2D HPs, the existence of Pb and I prompts a large spin-orbit coupling (SOC) which in the presence of structural inversion asymmetry boosts the spin-degeneracy of the CB and VB (Kepenekian et al., 2017, Kepenekian et al., 2015). It seems that 2D perovskites self-assemble into regular "quantum-well" structures that break the symmetry of the 3D system. It amplify the Rashba-effect and escalates their potential for opto-spintronic applications as observed by transient spectroscopic measurements (Zhai et al., 2017). The rate of charge carrier recombination and spin-coherence lifetimes of 2D RP perovskite single crystals, $PEA_2PbI_{4/}(MAPbI_3)_{n-1}$ (with n = 1, 2, 3, 4) has been reported in Ref (Chen et al., 2017). Charge carrier recombination rates are observed to be fastest for n = 1 due to the large exciton binding energy and the slowest for n = 2. Spin-coherence times at ambient temperature also demonstrate a non-monotonic layer thickness dependency with a growing spin-coherence lifetime, increases for n = 1 to n = 4, followed by a reduction in lifetime for n = 4 to ∞. For n=4, the longest coherence lifetime of ~7 ps was detected. Their observations are dependent on two contributions; a) Rashba-splitting increasing the spin-coherence lifetime for the n = ∞ to the smaller layered systems, b) phonon-scattering increases for smaller layers, reducing the spin-coherence lifetime. The interplay between these two contributions modifies the layer thickness dependency. Here they also measured the carrier dynamics by the exciton bleach kinetics and the spin-coherence dynamics by employing circularly polarized pump and probe pulses.

There from the above discussion we can extract that for device application a stable structure is the most feasible. Figure 6 describes the dimensional reduction, modified/engineered surface for a stable 2D HP device functionality. Remarkably, a report by G. Grancini *et al.* (Grancini et al., 2017) introduced a stable perovskite device by engineering $(HOOC(CH_2)_4NH_3)_2PbI_4/CH_3NH_3PbI_3$ 2D/3D perovskite junction. The unique gradually-organized 2D/3D multi-dimensional interface produces up to 12.9% efficiency in a carbon-based architecture, and ~14.6% in standard mesoporous SCs for 1 year. Their innovative, low-cost and stable architecture has supported the timely commercialization of PSCs. Thus, we propose that a detailed microstructural analysis of 2D HPs are the much needed approach to engineer new materials for energy research application specially SC. In this review we have tried to sum up the characteristics of 2D HPs reported / proposed to have stable architecture feasible for energy research application.

## 4 CONCLUSION

This review shows that for commercialisation of HP optoelectronic devices, a thorough in-depth characterisation of 2D HP materials are essential. Most unique feature essential for light absorbing layers of solar cell related to this type of material is their optimised dielectric constants and exciton binding energy, compared to conventional semiconductors, and it also include a rotational component connected with relaxation of molecular dipole. Although great efforts have been given to study various 2D HP materials, many challenges still there to utilize these materials in actual practical purposes. But, the fabricated devices using 2D HPs still facing degradation problem. More experimental and theoretical studies are needed in this regard to understand the degradation and improve the stability. On the other hand, there is a critical need of improved procedure of synthesis of single crystals of 2D HP with controlled thickness and large dimension. The above discussions highlights the correlation between the microscopic, crystallographic and molecular structure of 2D layered HP, and it is expected that these photo response properties of the material will be further corroborated towards developing flexible functional and tuneable optoelectronics device in future.

## 5    CONFLICT OF INTEREST

The authors declare that the review work was conducted in the absence of any commercial or financial connections that could be construed as a potential conflict of interest.

## 6    AUTHOR CONTRIBUTIONS

SB and GKC designed and wrote manuscript. PP supervised this work and revised the manuscript. All authors contributed to the article and agreed on the submitted version.

## 7    FUNDING

This work was supported by the Faculty Research Grant [NITC/DEAN(R&C)/FRG/2018-19/3], NIT Calicut and ANERT [ANERT-TECH/11/2019-S(NEP)], Govt. of Kerala, Kerala.

**FIGURE CAPTION:**

Figure. 1. A) Upper panel: The conventional crystallographic structure of cubic and orthorhombic CsPbBr3 perosvskite phases along with RP phases (for different values of n=1 and n=2). Lower panel: Unit cells and their projections along the [001] direction.

(B) Atomic structures of the domains (plausible RP) and the domain boundaries (between the phases of conventional perovskite and RP) and the comparison between the experimental and simulated STEM image.

Pure Cs and mixed Pb-Br cation columns are present along the [001] projected direction in conventional 3D structures of perovskite phases, as compared to the all mixed Cs-Pb-Br cation column in RP phases along the same projected direction. The figure is reprinted with permission from Y. Yu *et al. (*2017) copywrite 2017, American chemical society.

Figure 2. (A) Projection schematic of the 3D HP (Color codes are following: Red, describes a corner shared inorganic network of octahedral metal halide; Blue & Black, describes the interstitial position of the organic cations; Green, emphasizes the size restriction on cation).

(B) The projection of the 2D HP along the c-axis (Color codes are following: Blue & Black, represents the organic alteration; Red, represents the corner shared inorganic layers; Green, represents the restriction on area cross-section of organic cation not on the length). d= inorganic layer thickness and L=organic layer thickness for n=1.

(C) The energy band diagram of 2D HP structures ($E_g$ and $E_{exc}$ denote the electronic and optical bandgap). Blue color represents the $E_{exc}$.

The figure is reprinted with permission from D. B. Straus and C. R. Kagan *et al. (*2018) copywrite 2017, American chemical society.

Figure 3. The local conductivity and band alignment in 2D perovskite. (A) The device schematic comprising polycrystalline film of 3D perovskite with grain boundaries containing 2D perovskite.

(B) The layer by layer Tauc plots and UV photoelectron spectroscopic analysis of the device band structure.

The bare C-AFM images of FAPbI$_3$ (C, E) and with films of 2D perovskite (D, F) on ITO glass coated with SnO$_2$. 100 mV bias voltage was applied to get the images under the room light condition for (C, D) and low light intensity condition for (E, F) as provided by the setup. Topology of each film is depicted in the inset of the corresponding images. For (C, D) and (E, F), the scale bars are shown at the left and right side, respectively.

The figure is reprinted with permission from J.-W Lee *et al. (*2018) copywrite 2018, Nature publishing group.

Figure 4. Optical spectroscopy, PL (at 4 K blue square symbol and line) and absorption spectra (at 4 K red and 290 K green closed and opened circle) of the RP perovskite crystals with n =1 to 5.

(A) Scaling of experimentally obtained optical bandgap. (B) Photoluminescence spectra.

The figure is reprinted with permission from J. C. Blancon *et al. (*2018) copywrite 2018, Nature publishing group.

Figure 5. (A) The concentration gradient of quantum wells (QWs) throughout a film of layered perovskite is depicted here. The glass edge of the perovskite film is highly populated with the smallest QWs and the air edge is mostly dense with thickest QWs.

(B) Schematic diagram describing down-hill energy transfer cascade. Excitonic resonances of the quantum wells decrease with increase in thickness. The direction of electron transfer is same as that of energy transfer as compared to the holes which are moving towards the states of higher energy.

The figure is reprinted with permission from O. F. Williams *et al.* (2019) copywrite 2019, American chemical society.

Figure 6: Schematic representation of stability and functionality of 2D HP.

**(A)**

**(B)**

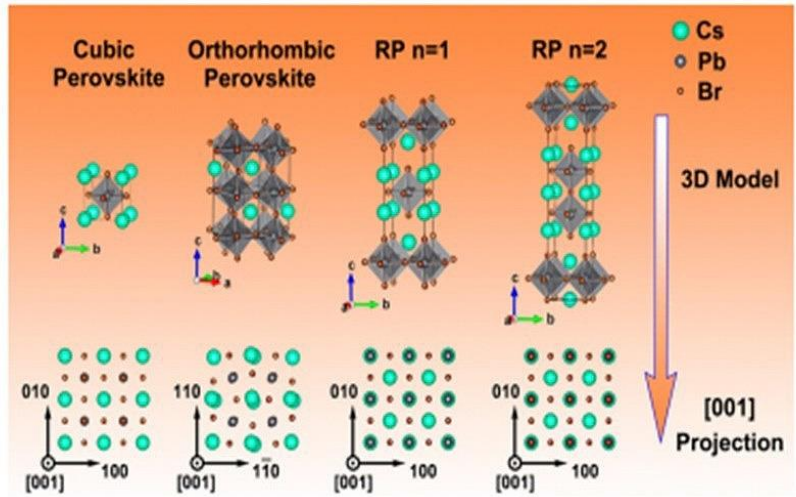

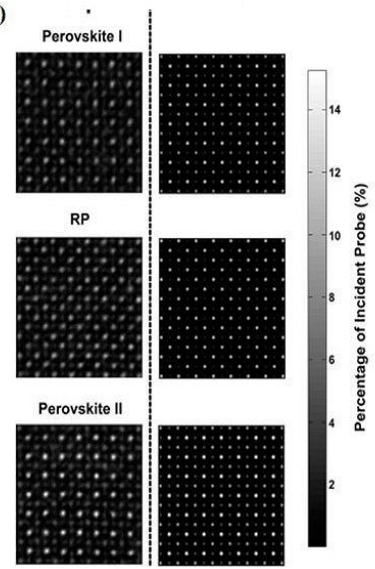

Figure 1

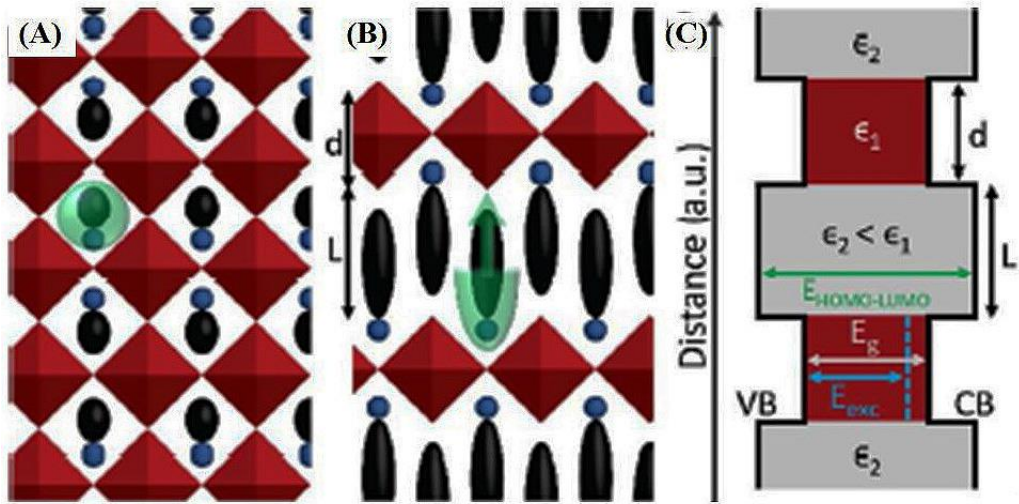

Figure 2

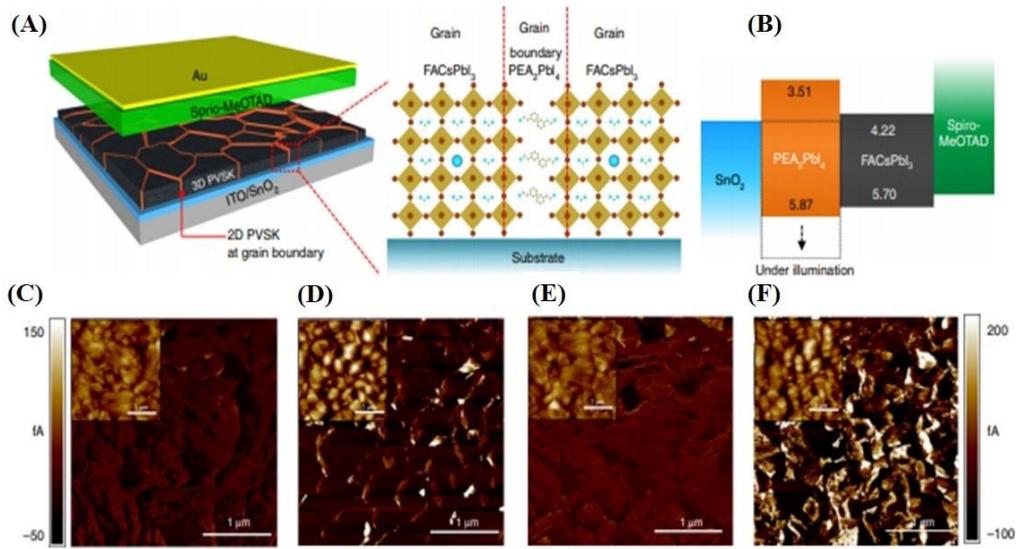

Figure 3

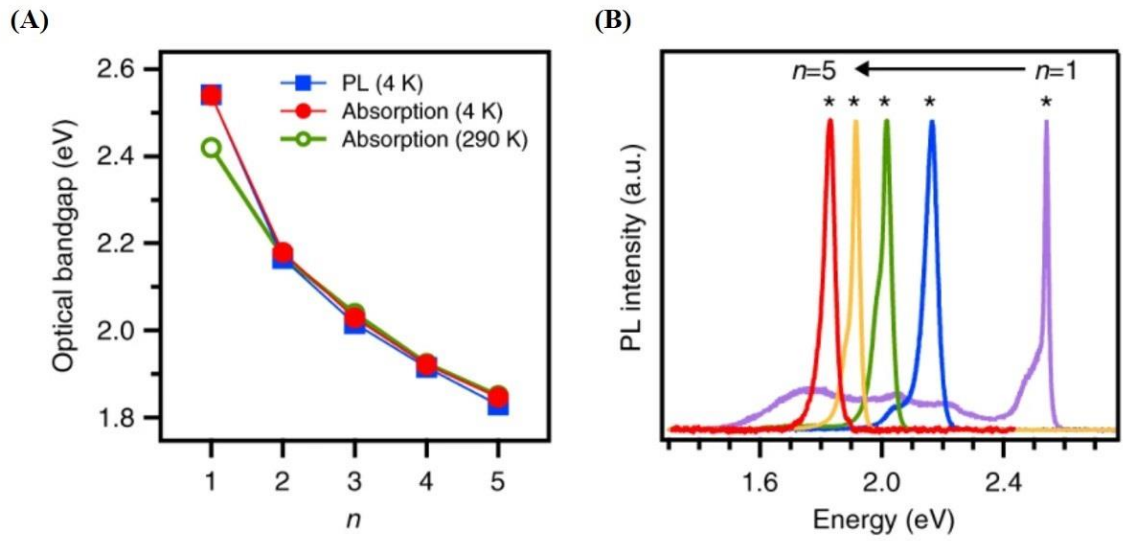

Figure 4

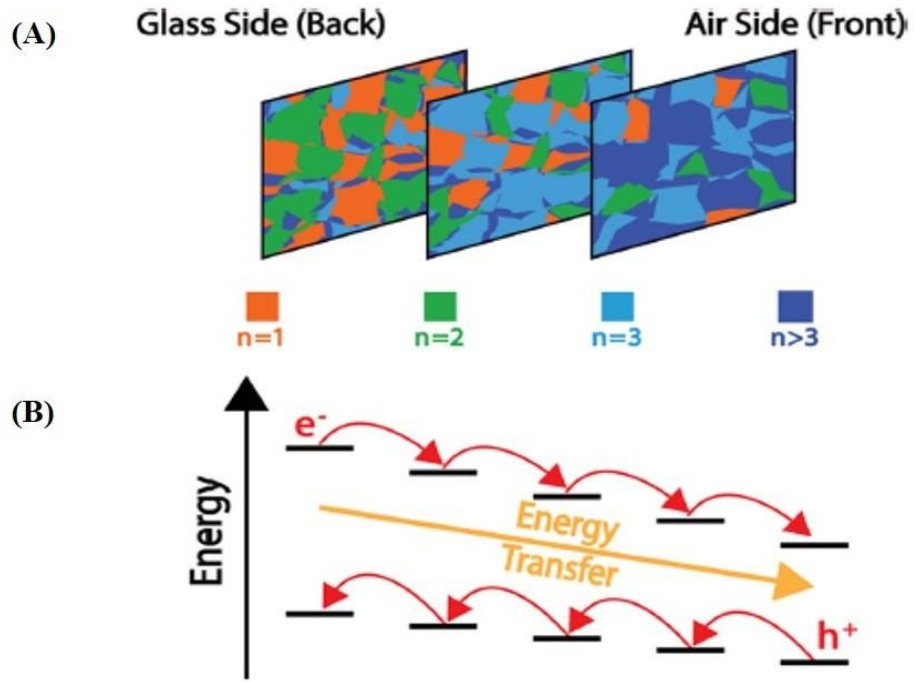

Figure 5

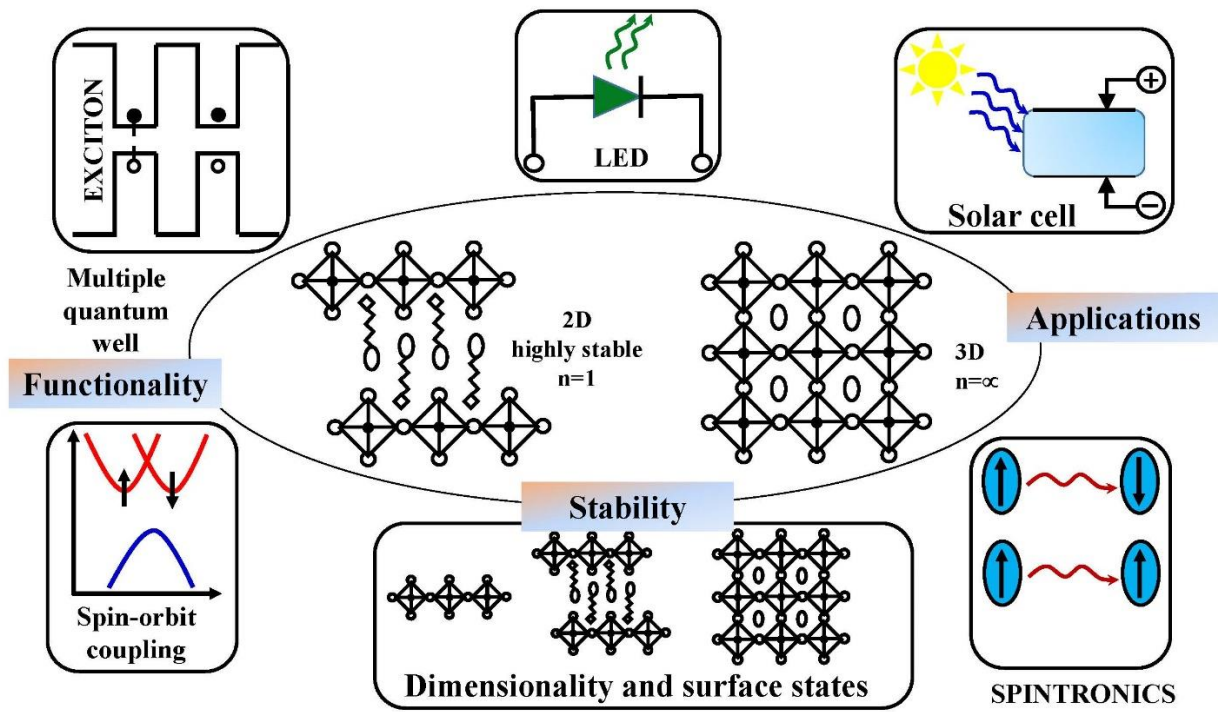

Figure 6